\documentclass[aps,twocolumn,showpacs,superscriptaddress,eqsecnum]{revtex4}
\usepackage{epsfig}
\usepackage{amsbsy}
\usepackage{amsmath}
\usepackage{amsfonts}

\begin{document}

\newcommand{\nn}{\nonumber}
\newcommand{\erf}[1]{Eq.\ (\ref{#1})}
\newcommand{\ip}[1]{\left\langle{#1}\right\rangle}
\newcommand{\bra}[1]{\langle{#1}|}
\newcommand{\ket}[1]{|{#1}\rangle}
\newcommand{\braket}[2]{\langle{#1}|{#2}\rangle}
\newcommand{\nb}{\overline{n}}
\newcommand{\half}{\tfrac 12}
\newcommand{\lr}[1]{\langle{#1}\rangle}

\title{Adaptive quantum measurements of a continuously varying phase}
\author{D.\ W.\ Berry}
\affiliation{Department of Physics and Centre for Advanced Computing 
--- Algorithms and Cryptography, Macquarie University, Sydney 2109, 
Australia}
\author{H.\ M.\ Wiseman}
\affiliation{Centre for Quantum Dynamics, School of Science, Griffith 
University, Nathan 4111, Australia}
\date{\today}

\begin{abstract}
We analyze the problem of quantum-limited estimation of a 
stochastically varying phase of a continuous beam (rather than a 
pulse) of the electromagnetic field. We consider both nonadaptive 
and adaptive measurements, and both dyne detection (using a local 
oscillator) and interferometric detection. We take the phase 
variation to be $\dot\varphi = \sqrt{\kappa}\xi(t)$, where $\xi(t)$ 
is $\delta$-correlated Gaussian noise. For a beam of power $P$, the 
important dimensionless parameter is $N=P/\hbar\omega\kappa$, the 
number of photons per coherence time. For the case of dyne 
detection, both continuous-wave (cw) coherent beams and cw  
(broadband) squeezed beams are considered. For a coherent beam a 
simple feedback scheme gives good results, with a phase variance 
$\simeq N^{-1/2}/2$. This is $\sqrt{2}$ times smaller than that achievable by 
nonadaptive (heterodyne) detection. For a squeezed beam 
a more accurate feedback 
scheme gives a variance scaling as $N^{-2/3}$, compared to $N^{-1/2}$ 
for heterodyne detection. For the case of 
interferometry only a coherent input into one port is considered. The 
locally optimal feedback scheme is identified, and it is shown to 
give a variance scaling as $N^{-1/2}$. It offers a significant 
improvement over nonadaptive interferometry only for $N$ of order 
unity.
\end{abstract}
\pacs{42.50.Dv, 42.50.Lc, 03.67.Hk}
\maketitle

\section{Introduction}
The phase of an electromagnetic field is not a quantity that can be 
directly measured. All phase-measurement schemes rely on measurement 
of some other quantity, which necessarily introduces an excess 
uncertainty in the phase estimate. The standard method of measuring 
the phase of a single mode is to combine it with a strong local-oscillator
field, which is detuned from the signal (so the phase 
changes linearly with respect to the signal phase). This is called 
the heterodyne scheme, and introduces an excess uncertainty scaling 
as $1/\nb$, where $\nb$ is the mean photon number. If the signal 
phase is known approximately beforehand, the introduced phase 
uncertainty can be reduced greatly by using a local-oscillator phase 
that is $\pi/2$ out of phase with the signal (homodyne measurements).

If there is no estimate for the phase available beforehand, it is 
still possible to reduce the excess phase uncertainty by adjusting 
the local-oscillator phase during the measurement so as to 
approximate a homodyne measurement \cite{Wis95c,semiclass,fullquan}. 
The mark II dyne measurements considered in Refs.\ \cite{semiclass} and 
\cite{fullquan} introduce an excess phase uncertainty scaling as 
$\nb^{-3/2}$. It is even possible to attain the theoretical limit, 
scaling as $\ln \nb/\nb^2$, using a more sophisticated feedback 
scheme \cite{unpub}.

The case of interferometry is quite similar. In interferometry we 
wish to measure the phase shift in one arm of an interferometer by 
counting photons in the output ports. If a phase shift varying 
linearly in time is introduced into the other arm (analogous to the 
heterodyne case), there is a large introduced phase variance scaling 
as $\bar{n}^{-1}$. On the other hand, if feedback is used to 
adjust the auxiliary phase shift adaptively, the introduced phase 
variance is greatly reduced \cite{short,long}.

These studies are all based on single-shot measurements, where the 
measurements are made on a single (one- or two-mode) pulse with
finite duration and a single fixed phase. In practice, if we wish to 
transmit information via a beam, a time-varying phase would be more 
convenient. A time-varying phase may also arise through random
fluctuations, and we may wish to keep track of the phase as well as
possible.

It is also possible to model a broadband signal that carries
information by random fluctuations. We therefore consider the case
of a phase subject to white noise in this paper.
We consider cw measurements for both dyne measurements and
interferometry. For the former, we consider both coherent beams and
broadband squeezed beams. For interferometry it is not clear if 
there is a cw analog to the optimal two-mode states derived 
in Refs.\ \cite{short,long}. Therefore, we consider only the case of a 
coherent input into one port.

\section{Adaptive Dyne Measurements on a Coherent Beam}

First, we will consider the case of cw dyne measurements on 
a single beam with a varying phase. It is simplest to consider a 
coherent beam with amplitude $\alpha=|\alpha|\exp[i\varphi(t)]$ 
having a constant magnitude, but varying phase. The magnitude is 
scaled so that $|\alpha|^{2}$ is the photon flux ($P/\hbar\omega$). 
As explained above, the phase is assumed to diffuse in time,
\begin{equation}
\label{varyphase}
\varphi (t+dt) = \varphi (t) + \sqrt{\kappa} dW'(t).
\end{equation}
Here $dW'$ is a Wiener increment satisfying $(dW')^{2}=dt$. The 
spectrum for the coherent beam is a Lorentzian of linewidth (full width at half
maximum) $\kappa$.

As in the single-shot case, a quadrature of the field is measured by 
combining the mode to be measured with a large-amplitude local-oscillator field
at a 50:50 beam splitter and measuring the outputs
with photodetectors. The photocurrent is then defined by
\begin{equation}
I(t) = \lim_{\delta t \to 0} \lim_{\beta \to \infty}
\frac{\delta N_+-\delta N_-}{\beta \delta t},
\end{equation}
where $\delta N_+$ and $\delta N_-$ are the outputs from the
photodetectors and $\beta$ is the local-oscillator amplitude.
For a continuous coherent beam this yields
\begin{equation}
\label{photo}
I(t)dt = 2{\rm Re} ( \alpha e^{-i\Phi(t)} )dt + dW(t),
\end{equation}
where $\Phi(t)$ is the phase of the local oscillator, and $dW(t)$ is a 
Wiener increment independent of $dW'(t)$.

In making adaptive phase measurements the phase of the local 
oscillator is usually taken to be $\Phi (t) = \hat \varphi (t) + 
\pi/2$, where $\hat \varphi (t)$ is some estimate of the system phase 
$\varphi(t)$ \cite{fn1}. With this, the signal becomes
\begin{equation}
I(t)dt = 2|\alpha| \sin \left[ \varphi (t) - \hat \varphi (t) 
\right] dt + dW(t).
\end{equation}

\subsection{Linear Approximation} \label{ss:linap}

Provided that the estimated system phase is sufficiently close to the 
actual system phase, we can make the linear approximation
\begin{equation}
\label{linearapprox}
I(t)dt = 2|\alpha|[ {\varphi (t) - \hat \varphi (t)} ]dt + 
dW(t).
\end{equation}
Rearranging this equation, we see that
\begin{equation}
\theta(t) =  {\hat \varphi (t) +  {I(t)}/{2\left| \alpha  \right|}} 
\end{equation}
is an unbiased estimator of $\varphi(t)$ based on the data obtained 
in the infinitesimal time interval 
$[t,t+dt)$. We will denote the best phase estimate based on all the data up 
to time $t$ by $\Theta(t)$. Note that this is the {\it best} phase 
estimate, in contrast to the phase estimate used in the feedback 
$\hat\varphi(t)$. The variance of each phase estimate $\theta(t)$ is 
given by
\begin{equation}
\label{bigvar}
\lr{ [ \theta(t) - \varphi (t) ]^2 } = \ip{\left( {\frac{dW(t)} 
{2|\alpha|dt}} \right)^2 } = \frac 1{4|\alpha|^2 dt}.
\end{equation}
Here the simple definition of the variance has been used, rather than 
the Holevo phase variance \cite{Hol84}
\begin{equation}
V_{H}(\Theta) = |\lr{e^{i\Theta}}|^{-2} - 1,
\end{equation}
as in Refs.\ \cite{semiclass,fullquan,unpub,short,long}. This is because we 
are using the linear approximation.

The noise in the estimate $\theta(t)$ is due entirely to the 
photocurrent noise, rather than the noise in the phase $\varphi$ 
itself. Since $dW(t)$ is independent of all previous noise, the 
updated best estimate $\Theta(t+dt)$ will be a weighted average of 
the instantaneous phase estimate $\theta(t)$ and the estimate from 
all the previous data $\Theta(t)$.

The equilibrium value of the variance of $\Theta(t)$, with all the 
individual phase estimates correctly weighted, will be denoted by 
$\Delta\Theta^2$. From \erf{varyphase}, 
after a time $dt$ the phase variance of $\Theta(t)$ 
with respect to the new system phase $\varphi(t+dt)$ will be 
$\Delta\Theta^2+\kappa dt$. The variance in the phase estimate from 
the latest time interval, $\theta(t)$, will be given by 
Eq.\ (\ref{bigvar}).

If we take a weighted average of $\Theta(t)$ and $\theta(t)$, then 
the contributions from each of the phase estimates from the 
individual time intervals should be correctly weighted, and the 
variance in the weighted average should be the equilibrium value, 
$\Delta\Theta^2$. This implies that
\begin{equation}
{\frac{1}{{\Delta\Theta ^2 + \kappa dt}} + 4|\alpha|^2 dt} = 
\frac{1}{\Delta\Theta ^2}.
\end{equation}
Solving for $\Delta\Theta^2$ gives $\Delta\Theta ^2 = 
\sqrt{\kappa}/{2|\alpha|}$.
If we define
\begin{equation}
N = |\alpha|^{2}/\kappa,
\end{equation}
the number of photons per coherence time (or photon flux divided by 
linewidth), we have
\begin{equation}
\Delta\Theta^{2} = {1}/{2\sqrt{N}}.
\end{equation}
This is the square root of the analogous result $1/4\bar{n}$ for a 
single-shot adaptive measurement on a coherent pulse of mean photon 
number $\bar{n}$.

Explicitly, the weighted average is
\begin{equation}
\Theta (t+dt) = \frac{( {4|\alpha|^2 dt} )\theta(t) + 
\Theta (t)/(\Delta\Theta ^2 + \kappa dt)}{1 /{\Delta\Theta ^2 }}.
\end{equation}
Solving this as a differential equation gives
\begin{equation}
\Theta(t) = 2|\alpha|\sqrt{\kappa}\int_{-\infty}^t {\theta 
(s)e^{2|\alpha|\sqrt{\kappa}(s-t)} ds}.
\end{equation}
Therefore, this method corresponds to a simple negative exponential 
scaling of the weighting.

We can also consider a more general negative exponential scaling 
given by
\begin{equation}
\label{negativeexp}
\Theta (t) = \chi \int_{-\infty}^t {\theta(s)e^{\chi (s-t)} ds}.
\end{equation}
Note that with this more general scaling, $\Theta(t)$ is no longer 
necessarily the best phase estimate. For most of the remainder of 
this paper, $\Theta(t)$ will be used in this more general sense, 
rather than as specifically the best phase estimate. The best phase 
estimate will be found by finding the optimum value of $\chi$. Taking 
the derivative of this expression with respect to time gives
\begin{equation}
\Theta (t+dt) = \chi dt\theta(t) + (1-\chi dt)\Theta(t).
\end{equation}
This means that this method is again a weighted average, except with 
a weighting that is not optimum. If we find the variance of both 
sides of this equation and solve for $\Delta\Theta^2$ we obtain
\begin{equation}
\label{sigma2}
\Delta\Theta^2 = \frac{\chi}{8|\alpha|^2} + \frac{\kappa}{2\chi}.
\end{equation}
This equation has a minimum of 
$\Delta\Theta^2=\sqrt{\kappa}/2|\alpha|$ for 
$\chi=2|\alpha|\sqrt{\kappa}$, reproducing the result found more 
directly above.

\subsection{Exact treatment}
\label{exactcase}
The results of the previous section are all using the linear 
approximation (\ref{linearapprox}). Although this approximation is 
very useful for obtaining the asymptotic value of the variance, it 
does not directly tell us what to do in the exact case. In the exact 
case for single-shot measurements \cite{semiclass}, rather than 
averaging phase estimates from each time interval, 
we determine $A_v$ and $B_v$, defined (for 
scaled time $v \in [0,1]$) as
\begin{equation}
A_v = \int_0^v {e^{i\Phi } I(u)du}, ~~~ B_v = -\int_0^v {e^{2i\Phi } 
du},
\end{equation}
and obtain the phase estimate from
\begin{equation}
\label{phaseest}
\Theta (v) = \arg \left( vA_v + B_v A_v^* \right).
\end{equation}
The intermediate phase estimate in the simplest (mark II) case 
\cite{semiclass} was
\begin{equation}
\label{intermphaseest}
\hat\varphi (v) = \arg A_v.
\end{equation}

We seek cw analogues of these formulas, that should reproduce 
the above linearized results in the appropriate (large $N$) regime. 
Guided by Sec.\ \ref{ss:linap}, we replace the definitions of $A_v$ 
and $B_v$ by
\begin{align}
\label{defineA}
A_t &= \int_{-\infty}^t e^{\chi(u-t)} e^{i\Phi } I\left( u \right)du ,
\\
B_t &= - \int_{-\infty}^t e^{\chi(u-t)}e^{2i\Phi} du, \label{defineB}
\end{align}
and continue to use $\arg A_t$ as the intermediate phase estimate 
$\hat\varphi(t)$. We will not consider any better intermediate phase 
estimates here, as these only give very small improvements over the 
mark II case for coherent states.

To find a formula for $\Theta(t)$, we can use a similar approach to 
that used in Ref.\ \cite{semiclass}. Let us ignore the variation of 
the system phase in \erf{defineA}. 
 Since we expect from Sec.\ \ref{ss:linap} that for 
large $N$ the optimal $\chi$ is ${\rm O}(|\alpha|\sqrt{\kappa}) = 
{\rm O}(\kappa\sqrt{N}) \gg \kappa$, this is a reasonable approximation. 
Then we find
\begin{equation}
\label{expandA}
A_t = {\alpha}/{\chi} - \alpha^* B_t + i\sigma_t,
\end{equation}
where
\begin{equation}
\sigma_t = \int_{-\infty}^t e^{\chi(u-t)} e^{i\left(\Phi-\pi/2 
\right)} dW(u).
\end{equation}
Equation (\ref{expandA}) is analogous to the corresponding result 
\cite{semiclass} for the case of single-shot measurements, except 
with $v$ replaced with $1/\chi$. Note that from this derivation it 
naturally emerges that we should use the same exponential in the 
integrand for $B_t$ as for $A_t$. From Eq.\ (\ref{expandA}) it can be 
shown that
\begin{equation}
A_t + \chi B_t A_t^* = \alpha (1/\chi- \chi |B_t|^2 ) +
i\sigma_t - i\chi B_t \sigma_t^*.
\end{equation}
Taking the expectation value gives
\begin{equation}
\ip{A_t + \chi B_t A_t^* } \approx \alpha ( 1/\chi - \chi |B_t|^2 ).
\end{equation}
If the local oscillator phase is independent of the photocurrent 
record, then this is exact. In the case of feedback, $B_t$ may be 
correlated with $\sigma_t$, but this result should still be 
approximately true. Therefore, the phase estimate that will be used 
here is
\begin{equation}
\Theta(t) = \arg (A_t+\chi B_tA_t^*).
\end{equation}
Similarly to the single-shot case \cite{unpub}, we will define the 
variable $C_t = A_t + \chi B_t A_t^*$, so $\Theta(t)=\arg C_t$. The 
above derivation is not exact if the system phase is not constant; 
however, $\arg C_t$ should still be a good estimator for the phase in 
the semiclassical limit.

A differential equation for the feedback phase can be determined in a 
similar way as in Ref.\ \cite{semiclass}. Using Eq.\ (\ref{defineA}), 
we can determine the increment in $A_t$,
\begin{equation}
dA_t = e^{i\Phi} I(t)dt - \chi A_t dt.
\end{equation}
Taking the local oscillator phase to be $\Phi(t)=\arg A_t+\pi/2$, we 
find that
\begin{equation}
dA_t  = i\frac{A_t} {|A_t|}I(t)dt - \chi A_t dt,
\end{equation}
so the magnitude of $A_t$ varies as
\begin{align}
d|A_t|^2 &= A_t^* \left( dA_t \right) + \left( dA_t^* \right)A_t + 
\left( dA_t^* \right)\left( dA_t \right) \nn \\
&= \left(1 - 2\chi |A_t|^2 \right)dt.
\end{align}
Thus $|A_t|$ increases up to an equilibrium value given by 
$|A_t|^2=1/2\chi$.

Using this result, the increment in the feedback phase in the steady 
state is
\begin{align}
d\Phi(t) &= {\rm Im} [ d\ln A_t ] \nn \\
&= {\rm Im}\left[\frac{dA_t}{A_t}-\frac{\left(dA_t\right)^2} 
{2A_t^2} \right] \nn \\
&= \frac{I(t)dt}{|A_t|} = \sqrt{2\chi} I(t)dt.
\end{align}
Therefore, the feedback phase just changes linearly with the signal, 
with constant coefficient (rather than a time-dependent coefficient 
as in the pulsed case \cite{semiclass}).

Using this result gives the stochastic differential equation for the 
phase estimate $\hat \varphi (t)$ as
\begin{equation}
d\hat\varphi(t) = \sqrt{2\chi} \{2|\alpha|\sin [\varphi(t) - \hat 
\varphi (t) ]dt + dW(t) \} .
\end{equation}
Making a linear approximation gives
\begin{equation}
d\hat \varphi(t) = \sqrt{2\chi} \{ 2|\alpha|[\varphi(t) - \hat 
\varphi (t)]dt + dW(t) \} .
\end{equation}
Rearranging and integrating then gives the solution as
\begin{equation}
\hat \varphi (t) = \sqrt{2\chi} \int_{-\infty}^t e^{2|\alpha| 
\sqrt{2\chi}(u-t)} \left[ 2|\alpha|\varphi(u)du + dW(u) \right].
\end{equation}

If the phase is measured relative to the current system phase, then
\begin{equation}
\varphi(u) = -\sqrt{\kappa}\int_u^t dW'(s).
\end{equation}
To determine an expression for the phase estimate $\Theta(t)$, note 
that it can be simplified to
\begin{equation}
\Theta(t) = \hat \varphi(t) + \arg( 1+\chi e^{-2i\hat 
\varphi(t)} B_t ).
\end{equation}
Using \erf{defineB} and expanding the exponentials to 
first order gives
\begin{align}
\Theta(t) & \approx \hat \varphi(t) + \arg \left( 1-i\hat 
\varphi(t) +i\chi \int_{-\infty}^t e^{\chi (u-t)} \hat \varphi(u)du 
\right) \nn \\
& \approx \chi \int_{-\infty}^t \hat \varphi(u) e^{\chi (u-t)} du.
\label{linap2}
\end{align}
This demonstrates that the mark II phase estimate is approximately a 
weighted average of the intermediate phase estimates, just as in the 
pulsed case it is approximately an unweighted average 
\cite{semiclass}. Note also the similarity of this result to the 
result for the linear case (\ref{negativeexp}). Unfortunately the 
simple technique used in the linear case cannot be applied here. 
However, using the standard techniques of stochastic 
calculus, the expectation value $\langle \Theta^2(t) 
\rangle$ can be determined 
from \erf{linap2}, in a lengthy but straightforward calculation.
The 
result is exactly the same as that obtained using the linear 
approximation (\ref{sigma2}).

\section{Heterodyne Measurements on a Coherent Beam}
\label{hetero}
In order to determine how much of an improvement feedback gives for 
cw measurements, we will compare it with the case of 
cw heterodyne measurements. For heterodyne measurements on a 
pulsed coherent state, the introduced phase variance is equal to the 
intrinsic phase variance. This indicates that the first term in 
Eq.\ (\ref{sigma2}) should be double for the heterodyne case, so the 
phase variance is
\begin{equation}
\label{phvar79}
\ip{\Theta^2(t)} \approx \frac{\chi}{4|\alpha|^2} + 
\frac{\kappa}{2\chi}.
\end{equation}

We now show this more rigorously using a similar technique to that 
used in Ref.\ \cite{semiclass}. Expanding $A_t$ gives
\begin{equation}
A_t = \int_{-\infty}^t e^{\chi(u-t)} e^{i\Phi(u)} [ ( 
\alpha e^{-i\Phi} + \alpha^* e^{i\Phi} )du + dW(u) ].
\end{equation}
For the heterodyne case, the local oscillator phase $\Phi(t)$ varies 
very rapidly, so the second term above will be negligible. This means 
that $A_t$ simplifies to
\begin{equation}
A_t = |\alpha| \int_{-\infty}^t e^{\chi(u-t)} e^{i\varphi(u)} du + 
i\sigma_t.
\end{equation}
Since $B_t$ is also negligible, the phase estimate $\Theta(t)$ simplifies 
to $\Theta(t) = \arg A_t$. As above, the phase will be measured 
relative to the current system phase. In the limit $N\gg 1$, the 
system phase does not vary significantly during the time $1/\chi$, so 
we can take the linear approximation, giving
\begin{equation}
A_t \approx \frac{|\alpha|}{\chi} + i|\alpha| \int_{-\infty}^t 
e^{\chi(u-t)}\varphi(u)du + i\sigma_t .
\end{equation}
Using this, the phase estimate is
\begin{equation}
\Theta(t) \approx {\rm Im} \left[ i\chi \int_{-\infty}^t 
e^{\chi(u-t)} \varphi(u)du + i\chi \sigma_t/|\alpha| \right].
\end{equation}
Here the linear approximation has again been used. Further evaluating 
this gives
\begin{equation}
\Theta(t) = - \sqrt{\kappa}\chi \int_{-\infty}^t \!\!\! du\,e^{\chi(u-t)} 
\int_u^t dW'(s) + \frac{\chi}{2|\alpha|}(\sigma_t+\sigma_t^*).
\end{equation}
The variance is, therefore,
\begin{align}
\lr{\Theta^2(t)} &= \kappa \chi ^2 \left\langle \int_{-\infty}^t du_1 
\int_{-\infty}^t du_2 e^{\chi(u_1+u_2-2t)} \right. \nn \\
& ~~~\left. \times \int_{u_1}^t dW'(s_1) \int_{u_2}^t dW'(s_2) 
\right\rangle \nn \\ &~~~+ \frac{\chi ^2}{4|\alpha|^2} \ip{(\sigma_t 
+\sigma_t^*)^2}.
\end{align}
The first term here can be evaluated to give ${\kappa}/{2\chi}$. In 
addition, it is easy to show that $\lr{\sigma_t^2}\approx 0$ and 
$\lr{|\sigma_t|^2}=1/{2\chi}$. Using these results gives the variance 
as
\begin{equation}
\ip{\Theta^2(t)} =   \frac{\kappa}{2\chi} + \frac{\chi}{4|\alpha|^2}.
\end{equation}
This shows that Eq.\ (\ref{phvar79}) is correct. Using this result, 
the minimum variance is ${\sqrt{\kappa}}/{\sqrt 2 |\alpha|}$ for 
$\chi=\sqrt {2\kappa} |\alpha|$. In terms of $N$, this is 
$1/\sqrt{2N}$, which is $\sqrt 2$ times the minimum phase variance 
for the adaptive case.

\section{Results for Dyne Measurements on a Coherent Beam}

In order to verify the above analytical results, the equilibrium 
phase variance was determined numerically for a variety of 
parameters. Because we do not presuppose a value for $\chi$, there 
are two dimensionless parameters in our simulations,
\begin{equation} \label{dimless}
N = \frac{|\alpha|^{2}}{\kappa}, ~~~ {\rm X} = \frac{\chi}{|\alpha|^2}.
\end{equation}
From the above theory, the optimum value of ${\rm X}$ is $2/\sqrt{N}$ for 
the adaptive case and $\sqrt{2/N}$ for the heterodyne case.

The value of $N$ was varied from 1 up to $2.5 \times 10^{37}$. For 
each value of $N$, ${\rm X}$ was varied from a quarter to four times 
$2/\sqrt{N}$. Measuring time in units of $|\alpha|^{-2}$, the time 
steps used were $\Delta t = 1/10^3{\rm X}$. For these calculations 1024 
simultaneous integrations were performed and the variance was sampled 
repeatedly. The integrations were taken up to time $10/{\rm X}$, in order 
for the variance to reach its equilibrium value, then the variance 
was sampled at time intervals of $1/{\rm X}$ up until time $100/{\rm X}$.

\begin{figure}
\centering
\includegraphics[width=0.45\textwidth]{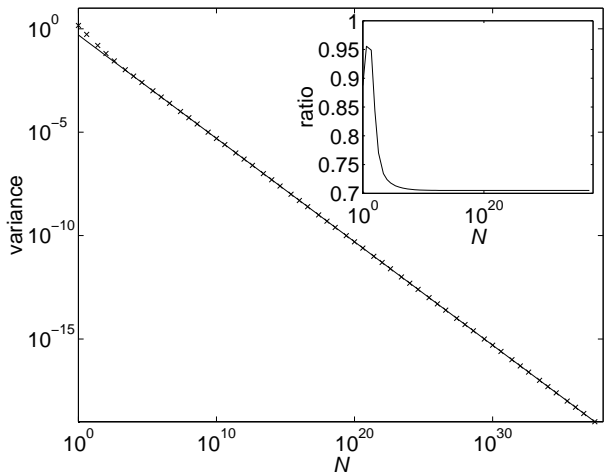}
\caption{The phase variance for cw adaptive measurements for 
${\rm X}=2/\sqrt N$. The numerical results are shown as crosses and the 
theoretical values of $1/\sqrt{2N}$ are shown as the continuous line. 
The inset shows the ratio of the minimum phase variance for 
cw adaptive measurements to the minimum phase variance for 
cw heterodyne phase measurements.}
\label{chi2k}
\end{figure}

The results for ${\rm X} = 2/\sqrt{N}$ are plotted in Fig.\ \ref{chi2k}. The 
variances for $N=1$ to $4\times 10^{12}$ are the Holevo variances, 
and for above $4\times 10^{12}$ are the standard variances. As can be 
seen, the results are very close to the theoretical values. To show 
the improvement over heterodyne measurements, the ratio of the 
minimum phase variance for adaptive measurements to the minimum phase 
variance for heterodyne measurements (with ${\rm X}=\sqrt {2/N}$) is 
plotted in the inset of Fig.\ \ref{chi2k}. The ratio is close to 1 for 
small $N$, but for larger $N$ the ratio gets closer and closer to 
$1/\sqrt 2$.

\begin{figure}
\centering
\includegraphics[width=0.45\textwidth]{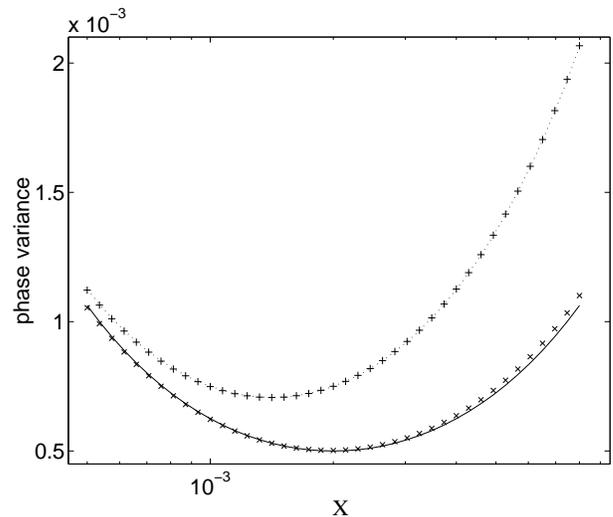}
\caption{The phase variance as a function of ${\rm X}$ for $N=10^6$. The 
numerical results for adaptive and heterodyne measurements are shown 
as the crosses and pluses, respectively, and the theoretical results 
for adaptive and heterodyne measurements are shown as the continuous 
line and dotted line, respectively.}
\label{bothpoor}
\end{figure}

Alternatively we can plot the phase variance as a function of ${\rm X}$ for 
fixed $N$. In Fig.\ \ref{bothpoor} we have shown the phase variance as 
a function of ${\rm X}$ for $N=10^6$, for adaptive and heterodyne 
measurements. The numerical results agree reasonably closely with the 
theoretical values, although there is a noticeable difference for 
adaptive measurements for the larger values of ${\rm X}$. Note that the 
minimum phase variance for adaptive measurements is at ${\rm X}=2/\sqrt N$, 
and the minimum phase variance for heterodyne measurements is larger 
and at a smaller value of ${\rm X}$. When the value of $N$ is increased 
further, the numerical results agree even more closely with the 
theoretical values.

\section{Adaptive Dyne Measurements on a Broadband Squeezed Beam}
\label{contsqz}

The above results show that the improvement offered by adaptive 
measurements over nonadaptive (heterodyne) measurements in the case 
of a coherent beam is only a factor of $1/\sqrt{2}$ reduction in the 
variance. This is similar to the single-shot case, where  a $1/2$ 
reduction was found for the coherent case. However, in the 
single-shot case a far more dramatic reduction is found for the case 
of a squeezed state. Motivated by this we now consider adaptive dyne 
measurements on a cw squeezed beam.

It is simplest to consider broadband squeezing. Physically, this 
could arise as the output of a driven parametric oscillator in the 
limit that the decay time of the cavity is much shorter than any 
other relevant timescales \cite{Gar91}. This results in the 
modification of the photocurrent from \erf{photo} to
\begin{align}
\label{photosqz}
I(t)dt&=2{\rm Re}(\alpha e^{-i\Phi(t)})dt + dW(t) \nn \\
&~~~ \times \sqrt{e^{-2r} 
\cos^2(\Phi-\phi_\zeta/2)+e^{2r}\sin^2(\Phi-\phi_\zeta/2)},
\end{align}
where $\alpha$ is the amplitude of the squeezed beam, and $r$ and 
$\phi_\zeta$ are the magnitude and phase of the squeezing, 
respectively. In this idealized limit the noise reduction via 
squeezing occurs by a reduction in the shot noise level, rather than 
an anticorrelation between the shot noise and the later coherent 
amplitude (as in the single-shot case).

For reduced phase uncertainty, the phase of the squeezing should be 
$\phi_\zeta=2\varphi + \pi$, where $\varphi$ is the system phase. If 
we are using feedback given by $\Phi = \hat\varphi + \pi/2$, where 
$\hat\varphi$ is an estimate of the phase, then the photocurrent can 
be expressed as
\begin{align}
I(t)dt&=2|\alpha| \sin(\varphi-\hat\varphi)dt \nn \\ &~~~
+dW(t)\sqrt{e^{-2r}\cos^2 
(\hat\varphi-\varphi)+e^{2r}\sin^2(\hat\varphi-\varphi)}.
\end{align}
It is clear that if the intermediate phase estimate used is very 
close to the system phase, then the factor multiplying $dW$ will be 
close to $e^{-r}$ and will be at a minimum. The better the 
intermediate phase estimate is, the smaller this multiplying factor 
will be. If the intermediate phase estimate is not perfect, it is 
clear that increasing the squeezing past a certain level will not 
reduce the multiplying factor. This is because the $e^{2r}$ term will 
start to dominate.

It is possible to estimate the optimum squeezing and the minimum 
phase variance using the linear 
approximation. In this approximation, the variance in the individual 
phase estimates $\theta(t)$ is  
\begin{equation}
[{e^{-2r}\cos^2(\hat\varphi-\varphi)+e^{2r}\sin^2(\hat\varphi-\varphi)}]/ 
{4|\alpha|^2 dt}.
\end{equation}
It is clear that the minimum phase variance (in this approximation) 
will be obtained when the best phase estimates are used for 
$\hat\varphi$. It is therefore reasonable to use the phase estimates 
$\Theta(t)$ for $\hat\varphi$, rather than $\arg A_{t}$ as in the 
coherent case. The values of $\Theta(t)$ will be the best phase 
estimates when the correct $\chi$ is used. As the variance of these 
estimates is $\Delta\Theta^2$, we obtain
\begin{equation}
\lr{e^{-2r}\cos^2(\hat\varphi-\varphi)+e^{2r}\sin^2(\hat\varphi-\varphi)} 
\approx e^{-2r} + e^{2r} \Delta\Theta^2.
\end{equation}
This approximation will be true for small phase variances and large 
squeezing. Following the same derivation as for the coherent 
case, the only difference is the multiplying factor, so we obtain
\begin{equation}
\label{simpler1}
\Delta\Theta^2 = \frac{\chi}{8|\alpha|^2}\left(e^{-2r} + e^{2r} 
\Delta\Theta^2\right)+\frac{\kappa}{2\chi}.
\end{equation}

This expression has two independent variables, $\chi$ and $r$, that 
can be varied in order to find the minimum phase variance. Taking the 
derivative of Eq.\ (\ref{simpler1}) with respect to $\chi$ and setting 
the result to zero gives
\begin{equation}
\label{opchi}
\chi = \frac{\kappa}{\Delta\Theta^2}.
\end{equation}
Substituting this into Eq.\ (\ref{simpler1}) gives
\begin{equation}
\label{simpler2}
\Delta\Theta^2 = \frac{\kappa}{4|\alpha|^2}\left( e^{2r} + 
\frac{e^{-2r}} {\Delta\Theta^2} \right).
\end{equation}
Taking the derivative of this with respect to $r$ and again setting 
the result equal to zero gives
\begin{equation}
\label{e2r}
e^{-4r} = \Delta\Theta^2.
\end{equation}
Substituting this back into Eq.\ (\ref{simpler2}) gives the phase 
variance as
\begin{equation}
\label{therval}
\Delta\Theta^2 = \left( \frac{\kappa}{2|\alpha|^{2}} \right)^{2/3} = 
\left(\frac{1}{2N}\right)^{2/3}.
\end{equation}

Thus we see that even for an arbitrarily squeezed beam, the best 
scaling we can obtain for the phase variance is $N^{-2/3}$, as 
compared to $N^{-1/2}$ for a coherent beam. This difference is less 
than for pulsed measurements, where the phase variance for the 
optimum squeezed states scales almost as $\nb^{-2}$, as compared to 
$\nb^{-1}$ for coherent states.

\section{Heterodyne Measurements on a Broadband Squeezed Beam}

In order to determine the phase variance for heterodyne measurements 
on a squeezed beam, we can simply perform the derivation of 
Sec.\ \ref{hetero}, except with the factor multiplying $dW$ from 
Eq.\ (\ref{photosqz}) included. This means that the variance will be
\begin{equation}
\lr{\Theta^2(t)} = \frac{\kappa}{2\chi} + 
\frac{\chi^2}{4|\alpha|^2}\lr{(\sigma_t +\sigma_t^*)^2} ,
\end{equation}
except with $\sigma_t$ modified to
\begin{align}
\sigma_t &= \int_{-\infty}^t e^{\chi(u-t)} e^{i\left(\Phi-\pi/2 
\right)} \nn \\
&~~~ \times \sqrt{e^{-2r} 
\sin^2(\Phi-\varphi)+e^{2r}\cos^2(\Phi-\varphi)} dW(u) .
\end{align}
Here we have used the assumption that the phase of the squeezing is 
$2\varphi+\pi$. Note that the derivation of Sec.\ \ref{hetero} takes 
the phase relative to the current system phase. This means that to a 
first approximation we may take $\varphi(u)=0$.

In order to determine the phase variance, we must determine the 
expectation values $\lr{|\sigma_t|^2}$ and $\lr{\sigma_t^2}$. We find
\begin{equation}
\lr{|\sigma_t|^2} = \int_{-\infty}^t e^{2\chi(u-t)} \left( e^{-2r} 
\sin^2\Phi+e^{2r}\cos^2\Phi \right) du .
\end{equation}
As the local oscillator phase $\Phi$ is varying rapidly in the heterodyne 
case, we may take the average values of $\sin^2$ and $\cos^2$, giving
\begin{equation}
\lr{|\sigma_t|^2} = \frac{\cosh(2r)}{2\chi}.
\end{equation}
Similarly, evaluating $\lr{\sigma_t^2}$ gives
\begin{equation}
\lr{\sigma_t^2} =  -\int_{-\infty}^t e^{2\chi(u-t)} e^{2i\Phi} 
\left( e^{-2r} \sin^2\Phi+e^{2r}\cos^2\Phi \right) du .
\end{equation}
Taking trigonometric averages as above gives
\begin{equation}
\lr{\sigma_t^2} = -\frac{\sinh(2r)}{4\chi} .
\end{equation}

Using these results we obtain the phase variance as
\begin{equation}
\lr{\Theta^2(t)} = \frac{\kappa}{2\chi} + 
\frac{\cosh(2r) - \half \sinh(2r)}{4|\alpha|^2/\chi}.
\end{equation}
This differs from the result for the coherent case by the multiplying 
term $ \cosh(2r) - \half \sinh(2r)$. This has a minimum of $\sqrt 
3/2$ for $r=\ln(3)/4$. Using this value, we obtain the minimum 
variance as $3^{1/4}\sqrt \kappa /(2 \alpha)$ for $\chi = 
2\sqrt\kappa |\alpha|/3^{1/4}$. Thus we find that the scaling is 
the same as for a coherent beam, and the multiplying factor 
is only about 7\% smaller. In contrast there is a factor of 
two difference in the single-shot case.

\section{Results for Dyne Measurements on a Broadband Squeezed Beam}

The results for the cw squeezed beam were obtained by a 
similar method as for the coherent case. 
Only variation in the variables $N$ and ${\rm X}$ of 
\erf{dimless} was considered, and time was measured in units of 
$|\alpha|^{-2}$. The step sizes used were $\Delta t = {1}/{10^3 {\rm X}}$. 
The integrations were taken up to time $30/{\rm X}$, then the variance was 
sampled every time step until time $130/{\rm X}$. The integration was 
performed using the photocurrent given in Eq.\ (\ref{photosqz}) with 
$\phi_{\zeta}=2\varphi+\pi$.

It was found that when  $\hat\varphi(t) = \arg C_t$ was used in the 
feedback, very poor results were obtained. This is a similar result 
to the case for single-shot measurements, where using $\arg C_v$ 
feedback results in large phase variances \cite{unpub}. This is 
because, when the intermediate phase estimates are extremely good, 
the results do not distinguish easily between the real system phase 
and the system phase plus $\pi$. This means that many of the results 
are out by $\pi$, resulting in a large overall phase variance.

In order to avoid this problem, rather than using $\arg C_t$ in the 
feedback, an intermediate phase estimate given by
\begin{equation}
\hat\varphi (t) = \arg (C_t^{1-\varepsilon} A_t^{\varepsilon})
\end{equation}
was used, with $\varepsilon$ constant. Note that this is the same as 
used to obtain phase measurements close to optimum in the single-shot 
case, except that there a time-varying $\varepsilon$ was used.

For each value of $N$ there are three variables that can be altered 
to minimize the phase variance: ${\rm X}$, $r$, and $\varepsilon$. It is not 
calculationally feasible to consider a range of values for all three 
variables. Instead, different values were tried systematically 
to find the minimum phase variance.

\begin{figure}
\centering
\includegraphics[width=0.45\textwidth]{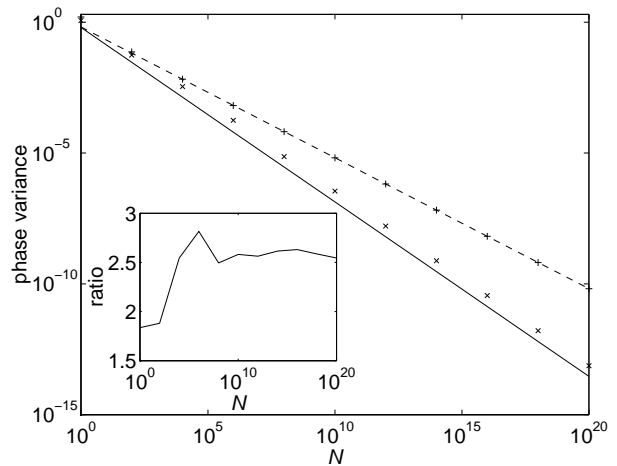}
\caption{The phase variance as a function of $N$ for a cw 
squeezed beam. The theoretical relations for adaptive and heterodyne
measurements are shown as the continuous line and dashed line respectively,
and the numerical results for adaptive and heterodyne measurements
are shown as the crosses and pluses respectively. The inset shows the
ratio of the numerically obtained phase variance to the theoretical value
as a function of $N$ for adaptive measurements.}
\label{mincontsqz}
\end{figure}

The minimum phase variances obtained by this method are plotted as a 
function of $N$ in Fig.\ \ref{mincontsqz}. The theoretical values 
given by Eq.\ (\ref{therval}) are also shown in this figure. The 
numerical results are higher than the theoretical values, but
 they have the same scaling with $N$, namely, $N^{-2/3}$. 
 If we plot the ratio of the 
numerical results to the theoretical values as in the inset of 
Fig.\ \ref{mincontsqz}, we find that for the largest values of $N$ the 
ratio levels off at about $2.6$. 

\begin{figure}
\centering
\includegraphics[width=0.45\textwidth]{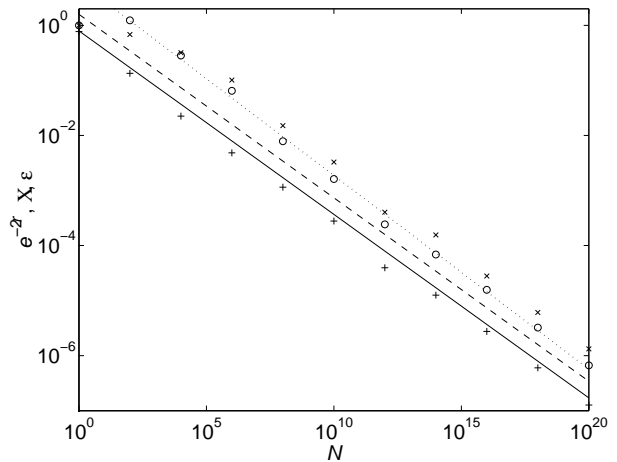}
\caption{The optimum values of $e^{-2r}$, ${\rm X}$, and $\varepsilon$ for 
measurements on a cw squeezed beam. The numerically found 
values of $e^{-2r}$ are plotted as crosses, and the theoretical 
expression as a continuous line. The numerically found values of ${\rm X}$ 
are plotted as pluses, and the theoretical expression as a dashed 
line. The numerically found values of $\varepsilon$ are plotted as 
circles, and the dotted line is the expression fitted to the data.}
\label{rccontsqz}
\end{figure}

Now note that, from Eqs.\ (\ref{e2r}) and (\ref{therval}), the optimum 
value of $e^{-2r}$ should be $(2N)^{-1/3}$. Similarly, from 
Eqs.\ (\ref{opchi}) and (\ref{therval}), the optimum value of ${\rm X}$ is 
$(N/4)^{-1/3}$. The numerically obtained optimum values of $e^{-2r}$ 
and ${\rm X}$, as well as these theoretical expressions, are plotted in 
Fig.\ \ref{rccontsqz}. Similarly to the case for the phase variance, 
the scaling is the same as theoretically predicted, but the scaling 
constants are different. For the case of $e^{-2r}$, the optimum 
values are about eight times those theoretically predicted, whereas the 
values of ${\rm X}$ are around a third of those theoretically predicted.

For the case of $\varepsilon$ there is no theoretical prediction for 
the optimum value. The numerically obtained values are shown in 
Fig.\ \ref{rccontsqz}, and as can be seen $\varepsilon$ decreases in a 
regular way with increasing $N$. A power law was fitted to these 
values (for $N>1$), and the power found was $-0.35\pm 0.01$. This is 
very similar to the $N^{-1/3}$ scaling found for $e^{-2r}$ and ${\rm X}$.

The results for heterodyne measurements are also shown in 
Fig.\ \ref{mincontsqz}. The results in this case agree very accurately 
with the theoretical prediction, within about $0.5\%$ for the larger 
values of $N$. Similarly the optimum values of $r$ and $\chi$ agree 
very accurately with those predicted above. The variance scales as 
$N^{-1/2}$, in contrast to the variance for adaptive measurements that scales 
as $N^{-2/3}$. This means that the improvement in using adaptive 
measurements scales as $N^{-1/6}$, which can be very large for large $N$.

\section{cw Interferometry}

\begin{figure}
\centering
\includegraphics[width=0.45\textwidth]{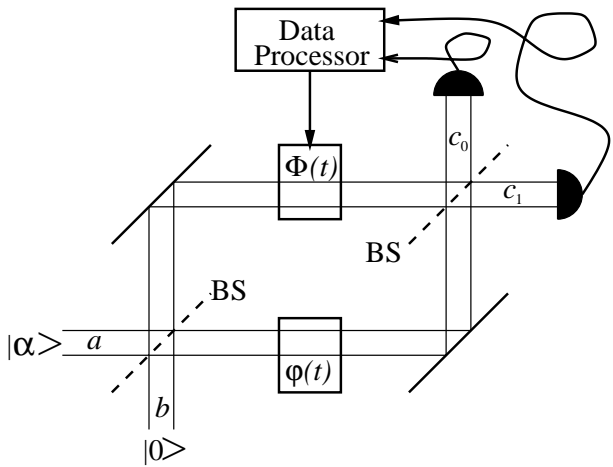}
\caption{The Mach-Zehnder interferometer, with the addition of a 
controllable phase $\Phi$ in one arm. The unknown phase to be 
estimated is $\varphi$. Both beam splitters are 50:50.}
\label{pict}
\end{figure}

Now we will turn from dyne measurement on a single beam to cw 
interferometric measurements. In this case we have a Mach-Zehnder 
interferometer (MZI), and are attempting to continuously track a 
stochastically varying phase in one arm, by controlling the phase in 
the other arm and detecting photons in the two output beams. This is 
shown in Fig.\ \ref{pict}. In this context it is not possible to 
consider nonclassical states of the type considered for the single-shot
case \cite{long}. Instead, for simplicity, we will restrict our 
consideration to the case where all photons enter through one port. 
This can be realized using coherent light, with $|\alpha|^{2}$ 
photons per unit time. Note that because this is an interferometric 
measurement rather than one using a local oscillator as a phase 
reference, the phase of $\alpha$ is irrelevant.

This case is essentially semiclassical, and the detections can be 
considered independently. Therefore, consider the state with a single 
photon incident on port $a$. The annihilation operators for the output
modes of the MZI, $\hat c_0$ and $\hat c_1$, are related to the
annihilation operators for the input modes, $\hat a$ and $\hat b$, by
\cite{long}
\begin{equation}
\hat{c}_{u}=\hat{a}\sin [ (\varphi-\Phi+u\pi)/2 ] +\hat{b}\cos 
[ (\varphi-\Phi+u\pi)/2 ],
\end{equation}
for $u\in\{0,1\}$. Hence the probability for detecting the photon
in detector $u$ is given by
\begin{equation}
\label{probdect} \sin^2 [ (\varphi-\Phi+u\pi)/2 ].
\end{equation}
Using Bayes' theorem, the probability distribution for the system 
phase after the detection is proportional to this probability times 
the initial probability distribution.

Denote the results for $m$ such detections by the string $n_{m} = 
u_{m}u_{m-1}\cdots u_{1}$. The probability distribution for the phase 
given $n_{m}$, $P(\varphi |n_m)$, can be expressed as
\begin{equation}
P(\varphi |n_m) = \sum_{k=-m}^m P_{m;k} (n_m)e^{ik\varphi}.
\end{equation}
In the absence of any phase variation, it can be shown from 
Eq.\ (\ref{probdect}) that the unnormalized coefficients 
$P'_{m;k}(n_m)$ can be determined by
\begin{align}
 P'_{m;k}\!(n_m\!) & \!=\! P_{m-1;k}\!(n_{m-1}\!) \!-\! \half
e^{-i(\Phi_m\!-u_m\pi)}\!P_{m-1;k-1}\!(n_{m-1}\!) \nn \\
&~~~ - \half e^{i(\Phi_m-u_m\pi)} 
P_{m-1;k+1} (n_{m-1}). \label{probrecurse}
\end{align}
The normalization condition on the probability distribution becomes 
$P_{m;0}(n_m)=1$. The normalized probability distribution can be 
obtained by simply dividing the coefficients obtained from 
Eq.\ (\ref{probrecurse}) by $P'_{m;0}(n_m)$.

Similarly to the case of dyne measurements, we will assume that the 
system phase diffuses with time as in Eq.\ (\ref{varyphase}). When the 
phase varies in time, the time between detections is important. For a 
photon flux of $|\alpha|^2$, the probability of a photodetection in 
time $dt$ is $|\alpha|^2dt$. The probability distribution for the 
time between detections is given by the exponential distribution
\begin{equation}
\label{timedist}
P_{E} (t)dt = |\alpha|^2e^{-|\alpha|^2t} dt.
\end{equation}
In the results that will be presented here, the time between 
detections, $\Delta t$, was determined according to this probability 
distribution.

Now in order to determine the effect of this phase diffusion on the 
probability distribution between detections, we must first consider 
the effect over some very small time interval $\delta t$. This is 
necessary because the probability distribution for the change in the 
system phase over time $\Delta t$ does not go to zero for $\Delta 
\varphi = \pm \pi$. This means that the probability distribution will 
not be exactly Gaussian, due to the overlap. In contrast, if we look 
at a very small time interval $\delta t$, the change in the phase 
will have a normal distribution with a variance of $\kappa \delta t$. 
Explicitly the probability distribution is
\begin{equation}
\label{phaseprobdist}
P_{G} (\Delta \varphi)d(\Delta \varphi) = 
\frac{1}{\sqrt{2\pi\kappa \delta t}} e^{-\Delta \varphi^2 /(2\kappa 
\delta t)} d(\Delta \varphi).
\end{equation}

The probability distribution for the phase after time $\delta t$ will 
be the convolution of the initial probability distribution with the 
Gaussian described by Eq.\ (\ref{phaseprobdist}). Evaluating this 
convolution gives
\begin{align}
P^{\delta t}(\varphi |n_m) &= \int_{-\pi }^\pi P(\varphi-\theta 
|n_m)P_{G} (\theta)d\theta \nn \\
&= \sum_{k=-m/2}^{m/2} P_{m;k} (n_m)e^{ik\varphi} \int_{-\pi}^\pi 
e^{-ik\theta} P_{G} (\theta)d\theta. \label{inthere}
\end{align}
As $\delta t$ is assumed to be small, $\kappa \delta t \ll 1$, and the 
integral in \erf{inthere} evaluates to $e^{-k^{2}\kappa \delta t/2}$.
The effect of the variation of the system phase on the probability 
distribution is, therefore,
\begin{equation}
\label{narrower}
P_{m;k}^{\delta t}(n_m) = P_{m;k}(n_m)e^{-k^2 \kappa \delta t/ 2} .
\end{equation}

In order to take account of the effect of the phase diffusion on the 
probability distribution over some significant time interval $\Delta 
t$, this time interval can be thought of as comprising $M$ small time intervals 
$\delta t$. Then we find that the coefficients are just multiplied by 
$M$ terms of $e^{-k^2 \kappa \delta t/2}$. This is equivalent to a 
single term of $e^{-k^2 \kappa \Delta t/2}$, which is very easy to 
implement.

As time passes the effect of Eq.\ (\ref{probrecurse}) is to broaden 
the distribution of probability coefficients in $k$, corresponding to 
a smaller variance in the phase distribution. In contrast, the 
Gaussian term in Eq.\ (\ref{narrower}) tends to narrow the 
distribution of probability coefficients, corresponding to a greater 
phase variance. The initially broad phase distribution narrows until 
an approximate equilibrium is reached, where the two effects cancel 
each other out.

In Ref.\ \cite{long} it was shown that the optimum phase estimate for
the single-shot case is
\begin{equation}
\Theta = \arg \lr {e^{i\varphi}} = \arg P_{m;-1} (n_m).
\end{equation}
It is easy to see that this phase estimate is optimal in the cw case
also. In addition we consider feedback that is equivalent to that
considered in the single-shot case in Ref.\ \cite{long}.
Rather than using an intermediate phase estimate as in the 
dyne case, we use the full power of Bayesian statistics to choose the 
feedback phase $\Phi$ so as to minimize the {\em expected} Holevo 
phase variance after the {\em next} detection. This is achieved by 
choosing $\Phi_{m}$ to minimize the value of \cite{long,thesis}
\begin{equation}
M(\Phi_{m}) = \sum_{u_m=0,1} \left| \int_{-\pi}^{\pi} P(n_m|\varphi) 
e^{i\varphi} d\varphi \right| .
\end{equation}
The values of $P(n_m|\varphi)$ can be obtained, except for a 
normalizing constant that is common to $u_m=0$ and 1, by using 
Eq.\ (\ref{probrecurse}). This means that we can express $M(\Phi_{m})$ 
as in Ref.\ \cite{long} with the parameters $a$, $b$, and $c$ given by
\begin{align}
a &= P_{m-1;-1}(n_{m-1}), \nn \\
b &= \half P_{m-1;-2}(n_{m-1}), \nn \\
c &= \half P_{m-1;0}(n_{m-1}).
\end{align}
These values of $a$, $b$, and $c$ can be used to determine the 
feedback phase as in Ref.\ \cite{long}.

The phase uncertainty at equilibrium can be estimated using a similar 
approach as was used for the single-mode case. Let us assume that the 
equilibrium variance in the best estimate for the system phase is 
$\Delta\Theta^2$. After time $\Delta t$, the variance in this phase 
estimate with respect to the new system phase, $\varphi(t+\Delta t)$, 
will be $\Delta\Theta^2+\kappa\Delta t$. In the equilibrium case this 
increase in the variance should, on average, be balanced by the 
decrease due to the next detection.

We now wish to estimate the equilibrium variance based on a weighted 
average with the previous best phase estimate, and a phase estimate 
from the new detection. If we use the actual variance for a phase 
estimate based on a single detection, then we do not get accurate 
results. This is because the variance for a single detection is 
large, so the weighted average does not accurately correspond to the 
exact theory. In order to make the theory based on weighted averages 
accurate, we need to assume an {\it effective} variance for the 
single detection, that is different from the actual variance.

In the case where there is no variation in the system phase, the 
phase variance after $n$ detections is approximately $1/n$ 
\cite{long}. It is clear that, if we assume that each detection has 
an effective variance of 1, then we will obtain the correct result. 
This is, in fact, equal to the variance as estimated using 
$\lr{2(1-\cos\varphi)}$ (this measure is used, for example, in 
Ref.\ \cite{measures}). Applying this to the case with a varying 
system phase gives
\begin{equation}
\frac{1}{\Delta\Theta^2  + \kappa \Delta t} + 1 = 
\frac{1}{\Delta\Theta^2}.
\end{equation}
Simplifying this to solve for $\Delta\Theta^2$, we find $\Delta\Theta 
^2 \approx \sqrt{\kappa \Delta t}$. On average, the time between 
detections is $1/|\alpha|^{2}$, so the approximate value of the 
variance should be
\begin{equation}
\Delta\Theta ^2 \approx \sqrt{\kappa/|\alpha|^{2}} = 1/\sqrt{N}.
\end{equation}

\section{Results for cw Interferometry}

In order to verify this theoretical result, the equilibrium phase 
variance was determined numerically for a variety of parameters. In 
this case there is only one dimensionless parameter, $N$. In the case 
of dyne measurements there was the additional parameter ${\rm X}$ 
describing how the latest results were weighted as compared to the 
previous results. In this case we do not have this parameter, as the 
phase estimates are not determined in that way.

The calculations were run for $10^5$ detections (or $2\times 10^5$ 
for the maximum value of $N$), and the phase error was sampled every 
detection after $10\sqrt{N}$ detections. This was done 100 times for 
each value of $N$. The equilibrium phase variance was determined in 
this way for the nearly optimum feedback scheme described above.
In addition we tested a nonadaptive measurement scheme with
\begin{equation}
\Phi_m = \Phi_0  + m \pi/\sqrt{N} ,
\end{equation}
where $\Phi_0$ is a random initial phase. When the value of $N$ was 1 
or less this was modified to
\begin{equation}
\Phi_m = \Phi_0 + m\pi/2,
\end{equation}
to prevent $\Phi_m$ being constant (modulo $\pi$). This is equivalent 
to the nonadaptive scheme in the single-shot case used in 
Ref.\ \cite{long}, and is analogous to heterodyne measurement. The 
reason for the factor of $1/\sqrt N$ is that the effective number of 
detections used for the phase estimate is $\sqrt{N}$. This follows 
from the fact that the phase variance is approximately $1/\sqrt{N}$.

A minor problem with cw adaptive measurements is that the 
number of probability coefficients $P_{m;k}(n_m)$ needed to determine 
the probability distribution for the phase rises indefinitely with 
the number of detections. The narrowing effect of the varying system 
phase, however, means that the probability coefficients fall 
approximately exponentially with $k$. The probability distribution 
can, therefore, be approximated very accurately by keeping only a 
certain number of coefficients. For the results presented here all 
probability coefficients with a magnitude above about $10^{-20}$ were 
used.

\begin{figure}
\centering
\includegraphics[width=0.45\textwidth]{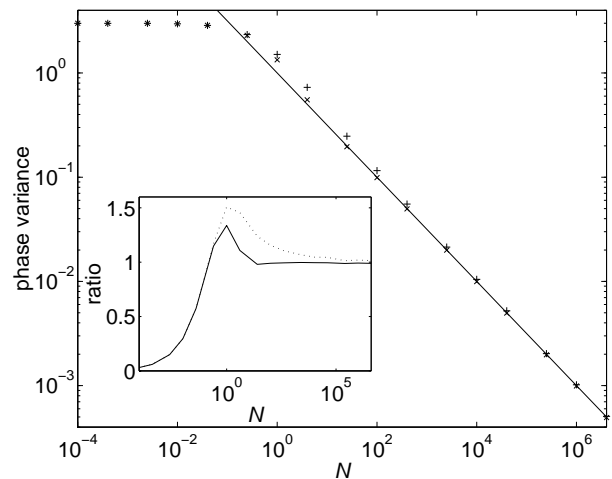}
\caption{The phase variance as a function of $N$. The numerical 
results for adaptive and nonadaptive measurements are shown as the 
crosses and pluses, respectively, and the theoretical values are shown 
as the continuous line. The inset shows the phase variance as a ratio 
to the theoretical value of $1/\sqrt N$. The results for adaptive and 
nonadaptive measurements are shown as the continuous line and dotted 
line, respectively.}
\label{runintlog}
\end{figure}

The Holevo phase variances for the two measurement schemes are 
plotted in Fig.\ \ref{runintlog}. 
As can be seen, the results for both cases are very close to the 
theoretical result of $1/\sqrt{N}$ 
for the larger values of $N$. For values of $N$ 
closer to 1 the results for the nonadaptive scheme are noticeably 
above the theoretical values. For small values of $N$ (less than 1), 
the variance converges to 3 for both the feedback schemes. This is 
what can be expected, as the system phase is randomized between 
detections. This means that the measurements are equivalent to phase 
measurements with a single photon, for which the Holevo phase 
variance is 3. The feedback has no effect, as there is no information 
on which to base it.

To see the differences more clearly, the phase variances are plotted 
as ratios to the theoretical values in the inset of 
Fig.\ \ref{runintlog}. The adaptive scheme gives phase variances that
are very close to, and slightly below, the theoretical
values for moderate values of $N$. In contrast, the results for
nonadaptive measurements are all above the theoretical values (for
$N\ge 1$). For small values of $N$ the variance for both schemes is
below $1/\sqrt{N}$, as the variance is converging to 3.

These results show that there will be a significant improvement in
using an adaptive scheme over a nonadaptive scheme only if the time
scale for the system phase variation is comparable to the time between
detections. This can be expected from the results for the single-shot
case with all photons in one port, where there was a significant
improvement in using an adaptive scheme only if the photon number was
small. The maximum improvement here is about 24\% for $N\approx 4$.

\section{Conclusions}
This study considered the problem of cw phase measurements, where the 
phase is being varied randomly in time and the aim is to follow this 
variation with the minimum possible excess uncertainty. We considered 
three different situations: dyne measurements on a  coherent 
beam, dyne measurements on a  (broadband) squeezed beam, 
and interferometric measurements using a coherent beam input. The relevant 
dimensionless parameter is $N$, the number of photons per coherence 
time (the characteristic time for the phase diffusion). Under optimum 
conditions, we found the analytical results, confirmed numerically, 
shown in Table \ref{table}. Previous results obtained 
for single-shot measurements on a 
pulse containing $n$, or $\bar{n}$ on average, photons are also shown for 
comparison. 

\begin{table*}
\caption{Scaling of phase variances for large photon numbers $N$ 
under various measurement conditions. For the cw (continuous-wave) 
cases, $N$ is the number of photons per 
coherence time. In the pulsed cases, $n$ ($\bar{n}$) is the (mean) photon 
number per pulse. Dyne measurements are those performed on a 
phase-shifted beam or pulse using a local oscillator, while MZI measurements
are of a phase shift in one arm of a Mach-Zehnder interferometer.
The two empty cells are those not treated in this study, and the question 
mark denotes a conjectured scaling. \label{table}}
\begin{tabular}{lcccc}
	\hline \hline
	  & ~Coherent, dyne~ & ~Squeezed, dyne~ & ~Coherent, MZI~ & ~Optimal, MZI~ \\
	\hline
	cw, adaptive & $N^{-1/2}/2$ & $O(N^{-2/3})$ & $N^{-1/2}$ & \\
	cw, nonadaptive & $N^{-1/2}/\sqrt{2}$ & $ N^{-1/2}\times 3^{1/4}/2$ & $N^{-1/2}$ & \\
	Pulsed, adaptive & $\nb^{-1}/4$ & $O(\ln \nb / \nb^{2})$ & 
	$n^{-1}$ & $O(\ln n / n^{2})?$ \\
	Pulsed, nonadaptive & $\nb^{-1}/2$ & $\nb^{-1}/4$ & $n^{-1}$ &
	$O(n^{-1})$ \\
	\hline \hline
\end{tabular}
\end{table*}

A number of regularities are evident from this table. 
With coherent light, the variance reduction offered 
by adaptive measurements is at most a multiplying factor. With 
nonclassical light, nonadaptive measurements scale in the same way as 
for coherent light, but adaptive measurements offer 
an improvement in the scaling. In all cases, the variance reduction 
(by a change in the prefactor or the scaling) is less in the cw case 
than in the pulsed case. This is because in order to obtain the best
phase estimate, as $N$ increases, the memory time for the estimate is 
reduced. This is needed to keep the 
contribution to the variance from the varying system phase (which 
increases with memory time) comparable 
with that from the quantum uncertainty (which decreases with memory 
time). This 
means that the effective number of photons used for the estimate is
the number per memory time, rather than the number per 
coherence time, $N$.

In the case of dyne measurements on a coherent beam, 
it was found that good results were obtained 
using a simple feedback phase ($\arg A_t$), similarly to mark II 
single-shot measurements \cite{semiclass}. In the cw case, the 
feedback simplifies to a form even simpler than for the single-shot 
case. Specifically, the feedback phase 
is simply adjusted proportional to the photocurrent. 
When the correct proportionality constant is selected, a minimum 
equilibrium phase variance is found that scales as 
$N^{-1/2}/2$. This is only $\sqrt{2}$ times smaller than 
the phase variance for heterodyne measurements.

For the case of dyne measurements on broadband squeezed states, the 
situation is considerably more complicated. The change in the phase 
cannot be taken to be proportional to the current, but rather is a functional 
with two parameters. With the degree of squeezing to be optimized as well, 
there are three parameters that 
must be varied to find the minimum phase variance. 
Nevertheless, it is still possible to obtain an analytic 
result that agrees with the numerical results in its scaling (although 
predicts the wrong multiplying factor). Specifically, it 
was found that the minimum phase variance varies as $N^{-2/3}$, 
compared to $N^{-1/2}$ for a coherent beam.
This contrasts with heterodyne measurements on 
broadband squeezed states, for which the minimum variance is only about 7\% 
below the corresponding result for a coherent beam. 

The case for interferometry is more difficult to treat, as it does 
not work with any simple feedback scheme. The feedback used was based 
on minimizing the expected variance after the next detection, 
similarly to the single-shot case. Despite this, it was found that it 
is possible to determine an approximate theory that agrees reasonably 
well with the numerical results for the case where a coherent beam enters 
one port of the interferometer. Similarly to the dyne case with a 
coherent state, the phase variance is proportional to $N^{-1/2}$. 
When a linearly changing feedback phase was used (analogous to the 
heterodyne scheme), it was found that the phase variance is above 
that for the adaptive feedback, but the difference is very small 
except for $N$ of order unity. This is as can be expected, as the 
difference is also very small for large photon numbers in the 
single-shot case.

In comparison with our previous pulsed results, 
the cw results obtained in this paper are probably more relevant to, and 
in some cases easier to implement in, a quantum-optics laboratory. 
This augurs well for future experimental verification.


\begin{thebibliography}{99}
\bibitem{fn1}
The only case where the feedback phase is not based on this (for dyne 
measurements) is for the corrections to the close-to-optimal 
measurements considered in Ref.\ \cite{unpub}.
\end{thebibliography}
\end{document}